\documentclass[twocolumn,prd,groupedaddress,nofootinbib,showpacs]
{revtex4}

\usepackage{latexsym}
\usepackage{amssymb}
\usepackage{graphicx}
\usepackage{dcolumn}
\usepackage{bm}
\input{epsf}
\begin{document}

\title{Global-String  and Vortex Superfluids in a Supersymmetric Scenario}

\author{
C.N. Ferreira $^{1}$
}
\email{crisnfer@pq.cnpq.br}

\author{J. A. Helay\"el-Neto $^2$}
\email{helayel@pq.cnpq.br}

\author{
W.G. Ney $^{1}$}
\email{wander@cefetcampos.br}

\affiliation{$^{1}$
N\'ucleo de Estudos em F\'{\i}sica, 
Centro Federal de Educa\c{c}\~ao Tecnol\'ogica de Campos\\
        Rua Dr. Siqueira, 273, Campos dos Goytacazes,\\
         Rio de Janeiro, Brazil, CEP 28030-130\\}

\affiliation{$^2$
Centro Brasileiro de Pesquisas F\'{\i}sicas, Rua Dr. Xavier Sigaud 150,
Urca,\\ Rio de Janeiro, Brazil, CEP 22290-180}

\date{\today}

\begin{abstract}
The main goal of this work is to investigate the possibility of finding the 
supersymmetric version of the U(1)-global string model which behaves as a vortex-superfluid. To describe the superfluid phase, we introduce a Lorentz-symmetry breaking background that, in an approach based on supersymmetry, leads to a discussion on the relation between the violation of Lorentz symmetry and explicit soft supersymmetry breakings. We also study the relation between the string configuration and the vortex-superfluid phase. In the framework we settle down in terms of superspace and superfields, we actually establish a duality between the vortex degrees of freedom and the component fields of the Kalb-Ramond superfield. We make also considerations about the fermionic excitations that may appear in connection with the vortex formation.

\end{abstract}

\pacs{12.60.Jv,11.27.+d}

\maketitle

\section{Introduction}

Stable vortex states may appear as an interesting manifestation of superfluidity. As one of the motivations, these vortices  have been observed in bosonic or fermionic diluted gases\cite{Madison2000,Shaeer99}. In the case of the bosonic vortices, the detection has been confirmed by analyzing the density variations in an expanding Bose-Einstein Condensate (BEC)\cite{Shaeer99,Cornell2002}. In Fermi systems, we do not expect significant density variations\cite{Nygaard} but, under certain conditions, the density variations may  be induced by the presence of one or more vortices that can be present in nuclear matter\cite{Yu}. The interior  of a neutron star, that is the only known system close to nuclear and neutron matter, constitutes  the appropriate  scenario for the vortex formation induced by the  rotational state of the star. 

There are important observations of astrophysical relevance that might be influenced by the presence of vortices in the interior of neutron stars; for instance, the pulsar glitches. The glitching events  represent a direct manifestation of the presence of superfluid vortices in the interior of the star, the triggering event being an unbalance between the hydrodynamical forces acting on the vortex and the force of interaction of the vortex with nuclei presence in the crust, pinning force \cite{Anderson,Pizzochero}; but, there are doubts about the value of the pinning force. 

One is related to the value of the energy gap in uniform neutron matter whereas the second problem is due to the very outlined way of treating vortex states in neutron matter. 
Global strings, which behave as a vortex superfluidity states, appear when a discrete symmetry is broken. 
These strings, as  the local strings  \cite{Vilenkin,Kibble1,Kibble2}, were most likely
produced during phase transitions \cite{Kibble}, and appear in some Grand-Unified Gauge Theories. They carry a large energy density \cite{Kibble1}. 
Both global and local strings  were mainly studied as  a possible mechanism for 
the seed density perturbation which has become a structure of large scale of the Universe we observe today \cite{Stebbins,Sato}. 

Nowadays, the approach considering cosmic string configurations has been revisited in connection 
with string theory\cite{Kibble,Polchinski,Majumdar} and the Wilkinson Microwave Anisotropy Probe (WMAP)\cite{Smoot}. The importance of this context is related to a possible measurement at the level of string theory, which  has supersymmetry (SUSY) as one of its main characteristics. SUSY is also related to cosmic strings in other contexts, where one contemplates the possibility that the boson-fermion symmetry was manifest in the early Universe, but it was broken approximately at the same time when these topological defects were formed. 

Many recent works investigate local strings by adopting a supersymmetric framework \cite{Morris,Davis,Ferreira02,Ferreira04,Ferreira05,Ferreira06}. In the context of the star formation \cite{Moulin}, SUSY appears as one of the most interesting mechanisms to describe cold dark matter\cite{Jungman,Masiero}. 
Both, local and global strings, are also important for their contribution to the gravitational radiation background\cite{Bennett}; in the case of the global symmetry, instead of radiating gravitationally, the dominant radiation mechanism for these strings is the emission of massless Nambu-Goldstone bosons \cite{Vilenkin1}. Global strings which behave as a vortex superfluidity states are connected with a Kalb-Ramond field. 
In some publications, it has been shown that, in  the low-energy regime, the effective action that presents the Kalb-Ramond fields, that also appear in string theory, provides
an accurate description of the dynamics of global strings\cite{Battye96}.

The Kalb-Ramond field\cite{KR,Lund} is an  antisymmetric tensor. This tensor, whenever interacting with a massive Higgs field, gives us a source. The system may have applications to superfluid helium and axion cosmology. A global vortex behaves as a superfluid if the Kalb-Ramond field breaks Lorentz symmetry in the background. The Kalb-Ramond fields in context of the topological defects can be studied in \cite{Braga1}, with SUSY framework \cite{Ferreira02,Ferreira06} and  associated with Lorentz-symmetry breaking can be studied in \cite{Ferreira04,Ferreira05}. For these implications, in this work, we analyze the equivalence of the vortex-superfluids to global strings in a supersymmetric context. This analogy is important to propose alternative models to be considered; the vortex stability and the  fermionic and bosonic behaviors of the matter can in the future enlighten us how to understand  the vortex states in fermionic  matter. 

The outline of this paper is as follows: in Section 2, we present some considerations about vortex superfluid models.

In Section II, we devote our attention to showing how the Lorentz-symmetry violation by the Kalb-Ramond background induces explicit SUSY breaking terms. In Section III, we focus on the general properties of the supersymmetric model for the vortex and treats some specific properties of the supersymmetric superfluid phase in the model we study. In this Section, we also carry out the superfield identifications at zero temperature and start a discussion on the fermionic excitations. Section IV Fermionic excitations are the main issue presented. In Section V, we propose a discussion on the non-zero temperature treatment of the model. Finally, in Section VI, we draw our General Conclusions.

\section{The implications of the Lorentz-violating background for soft SUSY breaking}

In this Section, we discuss  the
implications of the presence of the Lorentz-violating 
background for a soft SUSY breaking. The idea here is to
understand how the explicit soft SUSY breaking works to
yield mass to some of the field of the model.

We consider the possibility to get the scalar masses in SUSY theories by working with a 
Lorentz-symmetry violating background. This framework is important for a better understanding of the
relation between explicit SUSY breaking and Lorentz-symmetry violation. It is important to stress
that the problem of the SUSY breaking is a very important matter, related with the 
hiearchy problem and the mass constraints on the supersymmetric particles. In this section, let us 
start off with the following supersymmetric Lagrangian:

\begin{equation}
L_K =  \Phi^{\dagger}e^{4g{\cal G}} \Phi|_{\theta \theta \bar\theta \bar \theta}.\label{0}
\end{equation}
where the ingredient superfields of the model are: a chiral scalar supermultiplet, $\Phi(\phi,\chi,F)$, that contains a complex scalar field, $\phi $, a  spinor, $\chi_{a}$, and an auxiliary complex scalar field, $F$. 
The chiral scalar supermultiplet  $\Phi $ can be  $\theta $-expanded according to the following expression 

\begin{equation}
\Phi = e^{-i\theta \sigma^{\mu} \bar \theta \partial_{\mu}}[\phi(x) +
\sqrt{2}\theta^{a}\chi_a(x) + \theta^2F(x)], \label{sup1}
\end{equation}
${\cal G}$ is the Kalb-Ramond field-strength superfield defined in terms of the chiral spinor superfield as

\begin{equation}
{\cal G} = {1 \over 8}(D^a \Sigma_a - \bar D_{\dot a} \bar \Sigma^{\dot a})
\end{equation}
where

\begin{equation}
\begin{array}{lll}
\Sigma_a &=& \psi_a(x) + \theta^b \Omega_{ba}(x) + \theta^2\Big[\xi_a(x) 
+ i \sigma^{\mu}_{a \dot a}\partial_{\mu} \bar\psi^{\dot a}(x)\Big] \\
& &- i \theta \sigma^{\mu} \bar \theta \partial_{\mu} \bar \psi_a (x)- i \theta \sigma^{\mu} \bar \theta \theta^{\dot a} 
\partial_{\mu} \Omega_{\dot a a}(x)\\
& & - 
{1\over 4}\theta^2 \bar \theta^2 \Box \psi_a(x)
\end{array}
\end{equation}

The chirality condition for this field  is $ \bar D_{\dot a} \Sigma_a =0$. 

The Kalb-Ramond field accommodated in $\Omega_{\dot a b}(x)$ is given by

\begin{equation}
\Omega_{a b}= - \epsilon_{a b} \rho(x) + 
(\sigma^{\mu \nu})_{a b} {\cal B}_{\mu \nu}(x).
\end{equation}
with $\rho(x)$ and ${\cal B}_{\mu \nu}(x)$ being complex fields,

\begin{equation}
\begin{array}{ll}
\rho(x) = P(x) + i M(x),\\
{\cal B}_{\mu \nu}(x) = {1\over 4}\Big[B_{\mu \nu} - i \tilde B_{\mu \nu}(x)\Big]
\end{array}
\end{equation}
with
\begin{equation}
\tilde B_{\mu \nu}(x) = {1\over 2} \epsilon_{\mu \nu \alpha \beta}
 B^{\alpha \beta}(x)
\end{equation}

The components $P$ and $\psi_a$ are compensating fields and are not present in the $\theta$-expanssion of ${\cal G}$, as it shall be
explicitly given below.

The superfield ${\cal G}(M, \xi, \tilde G_{\mu} )$,
which plays a central role in connection 
with local vortices (\cite{Ferreira02}), accomodates the real scalar, M, the fermion $\xi$ and the dual of the Kalb-Ramond field strength $\tilde G_{\mu}$. It can be  $\theta$-expanded according to the following expression:

\begin{eqnarray}
{\cal G} &=&  -\frac{1}{2} M + \frac{i}{4} \theta^a \xi_a + \frac{i}{4} \bar
\theta^{\dot a} \bar \xi_{\dot a}+ \frac{1}{2} \theta \sigma^{\mu}_{a\dot
a} \bar \theta ^{\dot a} \tilde G_{\mu}  \nonumber\\
& &  + \frac{1}{8} \theta^a
\sigma^{\mu}_{a \dot a} \bar \theta^2 \partial_{\mu} \bar \xi^{\dot a} - \frac{1}{8} \theta^2 \sigma^{\mu}_{a \dot a} \bar \theta^{\dot
a}\partial_{\mu} \xi^a \nonumber \\
& & - \frac{1}{8} \theta^2 \bar \theta^2 \Box M;
\label{sup2}
\end{eqnarray}

Now, we have all the elements to illustrate how the  Lorentz-symmetry violation, signaled by the background of the Kalb-Ramond field,  is intimately connected to the appearance of explicit (soft) SUSY breaking terms.

We can notice that the 
superfield ${\cal G}$ carries only some degrees of freedom of $\Sigma_a $, 
the fermionic field $\psi _a $ does not appear, $\rho $ appears only 
through $M$, and as $\tilde G_{\mu}$ is related to the $2$-form, the Kalb-Ramond field, $B_{\mu \nu}$ 

\begin{equation}
\tilde G_{\mu} = \frac{1}{3 !}\epsilon_{\mu \nu \alpha \beta} G^{\nu \alpha \beta}.
\end{equation}
and

\begin{equation}
G_{\mu \nu \kappa} = \partial_{\mu}B_{\nu \kappa} + \partial_{\nu}B_{\kappa \mu} + \partial_{\kappa}B_{\mu \nu}
\end{equation}

\begin{eqnarray}
 L &=& \Big[ \partial_{\mu}  \phi^* \partial^{\mu} \phi + {i \over 4} \bar \chi \gamma^{\mu} \partial_{\mu} \chi + g^2|\phi|^2 \tilde G_{\mu} \tilde G^{\mu} \nonumber \\
& & + g \tilde G^{\mu}\Big( {1 \over 4} \bar \chi \gamma_{\mu} \chi\nonumber  - {i\over 2} \bar \phi \partial _{\mu} \phi  + {i \over 2 } \phi \partial_{\mu} \bar \phi  \Big)  \nonumber\\
& &   L_{_{Int}}\Big]\label{lg}
\end{eqnarray}
where in this discussion we consider ${\Phi} \rightarrow {\Phi} e^{g M}$. The Lagrangian ${L}_{_{Int}}$ is the interaction Lagrangian, and its explicit form is not important to  show the relation between the Lorentz and SUSY breakings.

Let 
us consider the split $G_{\mu \nu \lambda}$ as 

\begin{equation}
G^{\mu \nu \lambda} = G^{\mu \nu \lambda}_{(\rm self)}
+  G^{\mu \nu \lambda}_{(\rm ext)}.\label{breaking}
\end{equation}

The external Lorentz-symmetry breaking background is given by

\begin{equation}
G^{\mu \nu \lambda}_{(\rm ext)} = \sqrt{\rho}\epsilon^{0 i j k}= \sqrt{\rho}\epsilon^{i j k}. \label{external}
\end{equation}

The crucial point here is the justification of why the background value of $G^{\mu \nu \kappa}$ in (\ref{external}) yields an explicit soft breaking of SUSY. The whole idea here is that the background for $G^{\mu \nu \lambda}$ given in (\ref{external}) lies on a $\theta$-component of ${\cal G}$, actually, the $\theta \sigma^{\mu} \bar \theta \tilde G_{\mu} $
in (\ref{sup2}), which necessarily signals an explicit SUSY breaking. If the first component ( the $\theta$ -independent one) set up a non-trivial background then SUSY may not background, then SUSY may not be broken; however, whenever the background value sits on a non-trivial $\theta$-component, SUSY is necessarily explicitly broken down, and this is the case here.

The relevant bosonic part of the (\ref{lg}) important to analyzed the Lorentz-Breaking relation with SUSY breaking in the background is 

\begin{equation}
L =\partial_{\mu}  \phi^* \partial^{\mu} \phi + g^2|\phi|^2 \tilde G_{\mu} \tilde G^{\mu} 
-{ig\over 2} \tilde G^{\mu}\Big[ \phi^* \partial _{\mu} \phi - \phi \partial_{\mu} \phi^* \Big]  \label{Lorentz-breaking}
\end{equation}
By splitting the Kalb-Ramond fields as (\ref{breaking}) and by adopting the ansatz of a Lorentz-breaking background,
(\ref{external}) there emerges a mass term for the bosons,

\begin{equation}
L_{L-SUSY-B} =  g^2\rho |\phi|^2,\label{LB}
\end{equation}

Terms like that may appear as a result of spontaneous breaking of SUSY\cite{Nilles}. Soft explicit SUSY breaking terms are very important in connection with the physics derived from the Minimally Supersymmetric Standard Model (MSSM). In view of that, we try to stress here on the connection between a Lorentz-symmetry violating background and the appearance of explicit SUSY breaking terms.

\section{The Supersymmetric version for a Global Vortex and the Superfluid behavior }

In the present section, we study the supersymmetric framework setting up the general formalism that gives us the terms to construct the global vortex and study the superfluid behavior. The  action studied in the previous Section (\ref{0}) helps us in understanding the consequences of the Lorentz-breaking background in connection with SUSY. Actually, we have to see how the presence of a background yielding Lorentz symmetry violation also leads to an explicit SUSY breaking. In the present Section, let us adopt the action to study the vortex configuration as in the sequel:

\begin{eqnarray}
L_K & =&  \Phi^{\dagger}e^{4g{\cal G}} \Phi|_{\theta^2 \bar\theta^2 }  +
S^{\dagger} S|_{\theta^2  \bar\theta^2 } \nonumber\\
& & + W|_{\theta^2 } + \bar W|_{\bar \theta^2 }.\label{1}
\end{eqnarray}

These superfields satisfy a chirality constraint, given by the
condition $\bar D_{\dot a} \Phi = 0$ and $\bar D_{\dot a} {\cal S} = 0$. The superfield $\Phi$ is defined in  (\ref{sup1}) and ${\cal G}$ in (\ref{sup2}); the superfield $S$ has the same properties as $\Phi$ and  can be $\theta$- expanded according to  

\begin{equation}
{\cal S} = e^{-i\theta \sigma^{\mu} \bar \theta \partial_{\mu}}[S(x) +
\sqrt{2}\theta^{a}\zeta_a(x) + \theta^2H(x)]. \label{sup2.1}
\end{equation}
In the expression ({\ref{1}), $W$ is the superpotential whose general form is

\begin{equation}
W = a_i\Phi_i + b_{ij}\Phi_i \Phi_j + c_{i j k} \Phi_i \Phi_j \Phi_k .
\end{equation}

The scalar-field potential is given by

\begin{equation}
V = \sum_i \bar A_i A_i = \sum_i |{\partial W \over \partial \phi_i} |^2
\end{equation}  
where $A_i$ is the auxiliary component of the $\phi_i$-superfield. 

Let us study the possibility to obtain the supersymmetric version of the global vortex potential according to the model discussed in \cite{Davis89}. 

In the case of a global gauge symmetry, the chiral superfield $\Phi_i$ transforms as a phase under the $U(1)$-symmetry: 

\begin{equation}
\Phi'_i = e^{-i q_i \Lambda} \Phi_i\label{transformationschirality}
\end{equation} 
where $q_i$ are $U(1)$ global charges and $\Lambda$ is the rigid $U(1)$ rotation angle. The $q_i$ and $\Lambda$ are real constants.

It is possible to build up a  potential
with spontaneously broken gauge symmetry 
using three chiral superfields. For cosmic strings with a local gauge symmetry, one usually needs two charged fields $\phi_{\pm}$, with respective $U(1)$ 
charges $q_{\pm}$, and a neutral field, $\Phi_0$. This mechanism remains the same if we have a global transformation (\ref{transformationschirality}); in this case, the superpotential takes the form:

\begin{equation}
W(\Phi_i) = \mu \Phi_0\Big(\Phi_+\Phi_- - \eta^2\Big)
\end{equation}
In this approach, the neutral field $\Phi_0$ is important to give us the term responsible for the  mass of the scalar field of the theory, but, in a global vortex superfluid configuration, we have a Lorentz breaking background, as discussed in\cite{Davis89}, and we have proven, in the previous Section, that the Lorentz breaking introduces masses for the scalars; then, we can adopt a simpler potential, with one superfield $\Phi$, with charge $q_{\Phi}$, and another superfield, ${\cal S} $, with charge $q_{{\cal S}} $, satisfying the constraint $ q_{{\cal S}}  = -2 q_{\Phi}$, so that

\begin{equation}
W = h{\cal S} \Phi^2  \label{2}
\end{equation}

This form of the superpotential, in connection with the Lorentz-symmetry breaking (\ref{LB}) of the previous Section, leads to the Mexican hat configuration, responsible for the global vortex behavior that characterises superfluidity. 
 
The SUSY  transformations read as below:

\begin{equation}
\delta M = {i \over 2 } \bar \epsilon_{\dot a} \bar \xi^{\dot a} - {i \over 2} \epsilon^{a} \xi_a ,
\end{equation}

\begin{equation}
\delta \xi_a = 2 \sigma^{\mu}_{a \dot a} \bar \epsilon^{\dot a} \Big( \partial_{\mu} M - i \tilde G_{\mu}\Big)\label{transfer},
\end{equation}

\begin{equation}
\delta \tilde G^{\mu} = {i \over 2} \epsilon^b(\sigma^{\mu \nu})_b^{a}\partial_\nu \xi_a + {i \over 2} \bar \epsilon_{\dot b}(\bar \sigma^{\mu \nu})^{\dot b}_{\dot a} \partial_{\nu} \bar \xi^{\dot a}
\end{equation}

It is important to point out here that the soft supersymmetry breaking terms do not invalidate the supersymmetric transformations; actually, Lorentz symmetry and SUSY are broken down by the background, but they are both symmetries of the action. So, SUSY transformations as translations in superspace are not lost.

Global strings appear whenever a U(1) global symmetry 
is spontaneously broken. After the breaking of the U(1) symmetry, a massless Goldstone boson 
emerges that yields a long-range force.  The bosonic Lagrangian resulting from this supersymmetric model and that is relevant for the superfluid  can be written as

\begin{equation}
L_B = \partial_{\mu}\phi^{\dagger} \partial^{\mu} \phi + \partial_{\mu} S^{\dagger} \partial^{\mu} S + {g^2 \over 6} |\phi|^2 G_{\mu \nu \rho}G^{\mu \nu \rho} -  V' + \tilde G_{\mu}J^{\mu}. \label{5}
\end{equation}
for simplicity, we also adopt the redefinition $\Phi' \rightarrow \Phi e^{gM}$.  The current $j_{\mu}$ is given by

\begin{equation}
J_{\mu} = {-i g \over 2} \Big(\phi^* \partial_{\mu} \phi - \phi \partial_{\mu} \phi^* \Big)
\end{equation}

The bosonic potential $V'$ , that comes from  the eq.(\ref{2}), is given by

\begin{equation}
V' =  g^2 |\phi|^4  + 4g^2 |S|^2|\phi|^2. \label{6}
\end{equation}

We perform the background splitting (\ref{breaking}) of (\ref{fer6}), so that the full potential is

\begin{equation}
V =  h^2 |\phi|^4  - ( g^2\rho - 4h^2 |S|^2) |\phi|^2.\label{8}
\end{equation}

This potential shows us that the U(1)-breaking gives mass to the moduli field $|\phi |$ while the phase of the
scalar field remains massless.  In  the low-energy limit, the complex scalar  field of the Goldstone model 
can be represented as below:

\begin{equation}
\phi = \varphi(r)e^{i \alpha}\label{f},
\end{equation}
where $r$ is the radial coordinate. 
 The boundary conditions are given by

\begin{equation}
\begin{array}{ll}
\varphi(r) = 0 &\mbox{to }r =\delta\\
\varphi(r) = \eta &\mbox{to } r \rightarrow \infty .
\end{array}\label{cons}
\end{equation}

The configuration (\ref{f})-(\ref{cons}) is the same as the one of the local vortices, but the long-range interactions of global strings, happen due to their coupling to a massless Goldstone field, cause their dynamics to be substantially different from those of the local strings.
Spontaneous symmetry breaking requires that $\varphi(r)$ have mass and $\alpha $ be a massless Goldstone boson. This breaking  triggered by the soft SUSY breaking term we introduce, takes place whenever

\begin{equation}
( g^2\rho - 4h^2 |S|^2) > 0. \label{9}
\end{equation}

This relationship is crucial to ensure the stability of the potential, for it guarantees the vortex is formed around the right ground state. This justifies our claim, stated in the previous Section, on the importance of the term that breaks SUSY explicitly (and softly) for the stability of the global vortex. 
A solution to the vortex configuration exists if $g^2 \geq 4h^2 |S|^2 $. In this configuration, we consider the boundary conditions to $\varphi $, given by (\ref{cons}) and, for the S-field, we consider the ansatz $S = s(r)e^{i \Lambda}$. Outside the string, we consider the field $\langle S \rangle =0$.
The global vortex presents a
minimum roll and a central maximum characterizes the Mexican hat potential. By analyzing the potential minimum outside the string, with $\langle \phi \rangle = \eta$ and $\langle S \rangle =0$, we have $\eta = {g\rho \over h}$. The ansatz in the core of the string allows us to analyze it in comparison with the fermionic Yukawa potential that are the subject of the section IV. 


The  effective Lagrangian

\begin{eqnarray}
L_B &=&  \partial_{\mu}\varphi\partial^{\mu}\varphi + \varphi^2 \partial_{\mu} \alpha \partial^{\mu} \alpha  + \partial_{\mu} s\partial^{\mu} s + s^2 \partial_{\mu} \Lambda \partial^{\mu}\Lambda \nonumber \\
& +  & {g^2 \over 6} \varphi^2 G_{\mu \nu \rho}^{{\rm (self)}} G^{\mu \nu \rho}_{{\rm (self)}} +  {g^2 \over 3}\sqrt{\rho} \varepsilon^{i j k} \varphi^2 G_{ijk}^{{\rm (self)}}\nonumber \\
& + &{1\over 2} B_{\mu \nu}J^{\mu \nu}  -  V.
\end{eqnarray}

Now, let us write, the current $J_{\mu \nu}$ it in terms of the Kalb-Ramond field, according to the functional relation below:

\begin{eqnarray}
{\cal J} & = &\int \tilde G_{\mu}J^{\mu} d^4 x ={1 \over 2} \int \epsilon_{\mu \alpha \beta \gamma} \partial^{\alpha}B^{\beta \gamma} J^{\mu} d^4 x \nonumber \\ 
 & = &{1 \over 2}\int \epsilon_{\alpha \mu \beta \gamma} B^{\beta \gamma} \partial^{\alpha}J^{\mu} d^4 x = {1 \over 2} \int B_{\mu \nu}J^{\mu \nu} \label{current}
\end{eqnarray}
where

\begin{equation}
J_{\mu \nu} = {ig\over 2}\epsilon_{\mu \nu \alpha \beta} \partial^{\alpha}\Big(\phi^*\partial^{\beta}\phi - \phi \partial^{\beta} \phi^* \Big)
\end{equation}

 The configuration in the core of the string, where the  commutator is not zero, $[\partial_{\mu},\partial_{\nu}] \alpha \neq 0$, in the presence of a vortex. We can see this clearly by considering a straight vortex along the z- axis, the azimuthal angle, and integrating over a two-surface orthogonal to the string yields, $\int[\partial_x,\partial_y]\alpha dx dy =2 \pi,$ or $[\partial_x, \partial_y]\alpha = \frac{\delta(x)\delta(y)}{2\pi}$, then, in the  presence of the vortex the $\alpha $ is a multi-valued function of the coordinates and $J^{\mu \nu} \neq 0$ on the vortex core.

Outside the string core, $\phi$ can be represented as $\phi \sim \eta \exp(i \alpha(x))$ and $\langle S \rangle =0$; the effective Lagrangian for the Goldstone mode (in the presence of the global strings at large distances of the core, which are non-massive excitations) can be written as

\begin{eqnarray}
L& = &\eta^2 \partial_{\mu} \alpha \partial^{\mu} \alpha + {g^2 \eta^2 \over 6} G_{\mu \nu \rho}^{{\rm (self)}} G^{\mu \nu \rho}_{{\rm (self)}} \nonumber\\
& + &   {g^2 \eta^2\sqrt{\rho}  \over 3}\varepsilon^{i j k} G_{ijk}^{{\rm (self)}} + {1 \over 2} B_{\mu \nu}J^{\mu \nu} .
\end{eqnarray}

We use the fact that a real massless scalar field in four-dimensional Minkowski space is equivalent to a rank-2 anti-symmetric tensor , $B_{\mu \nu}$\cite{Lund,KR}; the nature of this equivalence in the case of the global strings can be found in \cite{Vilenkin1}. In SUSY, this duality property  can be understand by superfield identification

\begin{equation}
\Phi^{\dagger} \Phi \sim {\cal G}. \label{g}
\end{equation}

In fact, the left part that contains the vortex superfield gives us the 
term $\varphi^2\partial_{\mu}\alpha$ and the right side gives us a term related 
with the dual field,  $\tilde G_{\mu}$. The identification (\ref{g}) gives us other
contributions, related to the scalar field $M$:

\begin{equation}
\varphi^2 = 2\eta M\label{extra}
\end{equation}

The fermionic part is

\begin{equation}
\sqrt{2} \chi_a \varphi^* = - i\eta  \xi_a
\end{equation}

\begin{equation}
\sqrt{2} \bar \chi^{\dot a} \varphi =  i \eta \bar \xi^{\dot a}
\end{equation}
and the vortex identification part 

\begin{equation}
|\varphi^2| \sigma^{\mu} \partial_{\mu} \alpha + \bar \chi \chi   = \frac{1}{2} \eta \epsilon_{\mu \nu \lambda \rho}\sigma^{\mu}\partial^{\nu} B^{\lambda \rho}. \label{13}
\end{equation}

We can notice that the fermionic part modifies the usual vortex duality relation \cite{Davis89}. 
If we neglect the fermionic contribution, eq.(\ref{13}) can be written as

\begin{equation}
\varphi^2 \partial_{\mu} \alpha   = \frac{1}{2} \eta \epsilon_{\mu \nu \lambda \rho}\partial^{\nu} B^{\lambda \rho}. \label{14}
\end{equation}

The identification of the bosonic part given by (\ref{14}) has the same form as the  \cite{Davis89} for the global vortex configuration, but, with the supersymmetric invariance, we can always have a fermionic part.

The only remaining dynamical degree
of freedom is the scalar (Goldstone boson) field, $\alpha$.  In this approach, we have the action for the (global) static string:

\begin{widetext}

\begin{equation}
A = \int \frac{\beta}{6}\Big(G_{\mu \nu \beta}^{^{\rm (Self)}}G^{\mu \nu \beta}_{_{\rm (Self)}}+  2\sqrt{\rho} \varepsilon^{i j k} G_{ijk}^{{\rm (self)}} + {6\rho \over \beta}\Big) d^4x +{1 \over 2} \int B_{\mu \nu}J^{\mu \nu}d^4 x 
\end{equation}

\noindent
where $\beta = 1 + g^2 \eta^2 $.
\end{widetext}

At this point, it is advisable to remind that an explicit Lorentz-symmetry breaking, as stated above, may be rephrased in terms
of a softly explicit SUSY breaking term as the one we consider here \cite{Nos2008}. 
Now, let us study the solution at long distances compared to the string core, 
when the interaction of the vortex with the classical Goldstone-boson field is 
described by an effective Lagrangian. The stress tensor in the background 
considers a string at rest point in ${\bf \hat  u}$ direction we have

\begin{equation}
T^{00} =T^{ii} = \rho = p
\end{equation}

\begin{equation}
T^{0i} = \beta\sqrt{\rho} G^{0jk}_{_{self}} \epsilon^{i}_{\hspace{.1 true cm}jk}.
\end{equation}

The equation of the motion for the Kalb-Ramond field is

\begin{equation}
\partial_{\mu} G^{\mu \alpha \beta} = {1 \over \beta} J^{\alpha \beta}\label{current0}
\end{equation}

We obtain the solution

\begin{equation}
G^{0ij}_{_{\rm Self}} =  {g \sqrt{\rho} \over \beta h} {\hat u^i r^j - \hat u^j r^i \over r^2}
\end{equation}

It yields the stress tensor interaction part given by

\begin{equation}
T^{0i} = 2  {g  \over h} \rho{{(\bf \hat u}\times {\bf r})^i\over r^2}
\end{equation}

A single straight global string has a logarithmically divergent energy per unit of length. We can think that these strings could be ignored because they appear to be unphysical. However, following cosmological phase transitions, global strings may form loops with finite total energy or open strings with finite energy per horizon. An interesting application that some authors have been envisaging is the possibility that radiative decay of closed loops  be  connected with density fluctuation in the process of structure formation. This approach, considering the data basis, has been ruled out alone, but together with inflationary models and considering the noise of the experimental data, we can still consider them \cite{Smoot}. In the approach of  \cite{Davis89}, that is considered here, the vortex configuration is stable in the presence of the special background that breaks the Lorentz invariance\cite{Davis89}. The fact that the superfluid vortex is immersed in a Lorentz-noninvariant
fluid suggests that the correct model for a superfluid vortex involves the choice of a
special background. The relativistic force law for the response of a vortex
to the local field $G^{\mu \nu \rho}$ is  analogous to the Lorentz force law
in Electrodynamics. The external force due the background field interaction   is given as $F^{i}  = J_{j k} G^{j k  i} = \sqrt{\rho}\epsilon^{jk i} J_{j k} $\,. The bosonic part of the solution has the same form as in the non-supersymmetric model, but, in our construction,  the solution presents the explicit dependence on the parameters $h$ and $g$ and on the effects of the fermions.  In the supersymmetric version, the introduction of a Lorentz-symmetry violating background gives us  important implications on the fermionic background that we shall discuss in the next section, when we study the supersymmetric superfluid.

\section{The Analysis of the Fermions and the Superfluidity Behavior}

In a supersymmetric framework, besides the bosonic degrees of freedom, there are fermionic partners in the theory. In this section, let us analyse the behavior of the fermions that accompany the bosonic fields. The fermionic action can be written as:

\begin{equation}
L_F =  {i\over 2} \bar \chi \sigma^{\mu} \partial_{\mu}\chi  +{1 \over 2} B_{\mu\nu}{\cal J}^{\mu \nu} + L_{FKR} + L_{int}, \label{fer6}
\end{equation}
where ${\cal J}_{\mu \nu}$ is the fermionic current of the vorticity. The latter can also be expressed as follows:

\begin{equation}
{\cal J}_{\mu \nu} = {1\over 2}\epsilon_{\mu \nu \alpha \beta}\partial^{\alpha}\bar \chi \sigma^{\beta} \chi ,
\end{equation}
where

\begin{equation}
{\cal J}_{\mu} = {1\over 2}\bar \chi \sigma^{\mu} \chi .
\end{equation} 

The Lagrangian $L_{FKR}$ contains the fermionic Kalb-Ramond couplings and reads as below:

\begin{equation}
L_{FKR} = {g \over 2} \bar \chi \sigma^{\mu} \tilde G_{\mu} \chi . \label{FKR}
\end{equation} 

This Lagrangian (\ref{FKR}) amounts to a mass contribution given by the Lorentz-breaking parameter present in  (\ref{external}), that is,

\begin{equation}
L_{FKR}^{mass} = {\sqrt{\rho} g \over 2} \bar \chi \chi . \label{FKRmass}
\end{equation}

$L_{int}$ is the interacting Lagrangian, where we include the 
Yukawa terms which induce masses to the fermions that couple to the vortex. These Yukawa couplings are collected in:

\begin{equation}
Y = g\
\Big(2 \phi \zeta^a\chi_a + 2 \phi^* \bar \chi^a \bar \zeta_a + S \chi^a \chi_a + S^*\bar \chi^a \bar \chi_a \Big).\label{y}
\end{equation}

From  (\ref{y}) and (\ref{FKRmass}), there follows an interesting possibility. If we choose $\langle S \rangle =0$  to be zero in the core of the string, the mass (\ref{FKRmass}) does not appear in the core, the fermionic interaction term vanishes in the core and  the  fermions and  $\chi $ and $\xi $  become massless. In this case, where the fermions is not have mass, the fermionic zero-modes propagate with the speed of light in the z-direction and particles can be ejected from the vortex. Outside the vortex, these particles have masses induced by the $\phi$-interaction Yukawa term and by (\ref{FKRmass}), as induced by the Lorentz-symmetry breaking  \cite{Ferreira07}.

\section{Second-order phase transitions and the relation of the $S$-field with the temperature}

In this Section, let us study a physical interpretation of the field $S$. Up to now, we know that the field $S$
is important for the zero-modes. Now, let us study another interpretation, possibly related with second-order phase 
transitions. The potential (\ref{8}) can  represent a high-temperature effective potential\cite{Vilenkin1}, 
that can be written as

\begin{equation}
V(\phi,T) = m^2(T)|\phi|^2 + h^2 |\phi|^4\label{pottemp}
\end{equation}
where we identify $S $ with the temperature, $T$. We actually consider $|S|^2=T^2$, then

\begin{equation}
m^2(T) = h^2 (4T^2 - \eta^2).
\end{equation}

The term $m(T)$ is the mass for the $\phi$-field, whenever the state is symmetric, $\langle|\phi|\rangle = 0$.
This mass vanishes when $T = T_c$,  and

\begin{equation}
T_c = {\eta \over 2}.
\end{equation}

Another important case occurs for $T > T_c$; the effective mass $m^2(T)$ is positive and the minimum 
of $V$ is at $\phi =0$. The physical interpretation of this result is that the expectation value of $\phi $ vanishes. This means that the symmetry is restored at high temperature.  
The symmetric vacuum becomes unstable and $\phi$ develops a non-zero expectation value. 
Minimizing $V$, as in (\ref{pottemp}), we obtain, for $T < T_c$,

\begin{equation}
|\phi | = \sqrt{2}\Big(T_c^2 - T^2\Big)^{1/2}\label{phitemp}
\end{equation}  

An important realization of the second phase transition is the fact that $|\phi |$ grows
continuously from zero, as the temperature decreases from the critical temperature, $T_c$.

The cosmological point of view, when the supersymmetric Universe cools through the critical temperature, is that the field 
$\phi $ develops an expectation value of magnitude (\ref{phitemp}). The evolution of the phases $\alpha $ of 
$\phi$ and $\Lambda$ of $S$ with the temperature is not determined only by local physics;  their
values outside depends on random fluctuations and $\alpha $ and $\Lambda$ take different values in different regions of space during 
the evolution. But, since the free energy is minimized, these phases after the Universe 
expansion can become precedent sections with the temperature $T=0$. We can define the 
correlation length, $\Pi(t)$, to be the length scale above which the values of $\alpha$ and 
$\Lambda$ are uncorrelated.  The evolution of $\Pi(t)$ depends on details of the relaxation 
processes. Indeed, $\Pi(t)$ has to satisfy the causality bound. 
The correlation length cannot establish scales greater than the causal horizons  
related with the distance travelled by the light during the life-time of 
the Universe. For $T < T_c$, the scalar field develops an expectation value corresponding to some 
point in the manifold ${\cal M}$ of the minima of the effective potential $V$. We can see in (\ref{phitemp}) 
that the term that in our model is given by the soft SUSY breaking, 
was presented in the high temperature state, but, in the case  $T\gg T_c$,  SUSY
breaking can be neglected, and we can consider the Universe as being in a supersymmetric phase. To understand this fact, we need String Theory arguments, not contemplated in this work. 
This analysis only gives us knowledge about the vortex formation, and it is not able to
provide us with information on the Lorentz breaking. But, if we consider that, when the 
temperature becomes low vortex formation may take place, then Lorentz symmetry may be 
violated and there occurs a vortex-superfluid formation, as we have analysed throughout this paper.

\section{General Conclusions}

In this work, we have shown that it is possible to build up a string vortex by
modelling the vortex superfluid in a supersymmetric context.
We have analyzed the potential that gives us the correct string vortex
configuration, in zero temperature,
it presents a soft SUSY-breaking induced by the hidden sector. We have also
analyzed  the bosonic
aspects of the duality representation of the vortex. We also analysed the
physical interpretation of the extra field $S$ related to the presence of the
zero-mode and we show that we can relate it to the temperature. It is advisable
to comment here that the violation of Lorentz symmetry introduced is independent
of the soft SUSY explicit breaking terms. The latter has been considered to be
correct taking into account stability aspects of the potential. It would be
very interesting to eventually understand if the violation of Lorentz symmetry
and the explicit SUSY breaking could be related to one another. This would
render our proposal more interesting, in that we would be dealing with less
arbitrary parameters. Also, it would clarify the interplay between
Lorentz-symmetry violation (in the sense of particle transformations) and SUSY
explicit breaking that describes mass splittings among bosons and fermions that
belong to the same supermultiplet. This is an issue that could be investigated
better in a future work. The interesting phenomenological aspect of this discussion would be checking
whether properties like the masses of the SUSY particles, such as the photino
and the higgsino, would necessarily signal to some type of Lorentz-symmetry
breaking. In section V, we have proposed a discussion in a cosmological
evolution context, but, as already pointed out, these vortices also appear in
star cores. In the case of the neutron stars, as presented in the Introduction,
the force that dictates the vortex stability is induced by the nuclear matter;
but, we do not eliminate the possibility of the  presence of a Lorentz-symmetry
breaking to have an important role inside the star. Nothing guarantees that the
matter inside the star has a Lorentz-invariant behavior, because the high energy
envolves, in analogy with high energy $\gamma $-rays from extragalatic
sources\cite{Stecker}. Another point is if we consider that dark matter is
mostly composed by supersymmetric particles, the relation between the Lorentz
and SUSY breaking may become important to understand the parameters of 
the model. Then, the  possibility of the Lorentz-symmetry  breaking  in supersymmetric
matter becomes relevant for the dark matter stability around the stars
\cite{Jungman} and particles  can then be ejected out of these 
astrophysical structures. In this work, we do not have an application for these objects, but
we understand that our model can  be an alternative possibility to understand
some phenomena involving high energies. The next step is studying the fermionic
implication of the Lorentz-symmetry breaking Kalb-Ramond  background and how we
could find out a mechanism to justify its appearance, the relation between
Lorentz and SUSY breaking and the origin of the hidden sector represented by
soft breaking of global SUSY.

{\bf Acknowledgments:}

W. Bietenholz is acknowledged for a critical reading and for many helpful suggestions on an original manuscript.
The authors would also like to thank (CNPq-Brasil)  for the invaluable financial support.

\end{document}